\documentclass[10pt]{article}

\usepackage{graphicx}

\newcommand{\be}{\begin{equation}} 
\newcommand{\ee}{\end{equation}}
\newcommand{\bdm}{\begin{displaymath}} 
\newcommand{\edm}{\end{displaymath}}
\newcommand{\bea}{\begin{eqnarray}} 
\newcommand{\eea}{\end{eqnarray}}
\newcommand{\bse}{\small \begin{equation}} 
\newcommand{\ese}{\end{equation} \normalsize}
\newcommand{\bsea}{ \small \begin{eqnarray}} 
\newcommand{\esea}{\end{eqnarray} \normalsize}
\newcommand{\nn}{\nonumber \\} 
\newcommand{\non}{\nonumber}
\newcommand{\fig}{Fig.~\ref} 
\newcommand{\tab}{Table~\ref}
\newcommand{\sect}{Section~\ref}
\newcommand{\eqn}{Eq.~\ref}
\newcommand{\sgn}{\ensuremath{\mathrm{sgn}}}

\newcommand{\rmin}{\ensuremath{r_{\mathrm{min}}}}
\newcommand{\hmax}{\ensuremath{h_{\mathrm{max}}}}

\newlength{\diaght}
\newlength{\diagshift}
\newcommand{\mayeriva}{\raisebox{\diagshift}{\includegraphics[height=\diaght]{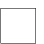}}}
\newcommand{\mayerivb}{\raisebox{\diagshift}{\includegraphics[height=\diaght]{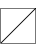}}}
\newcommand{\mayerivc}{\raisebox{\diagshift}{\includegraphics[height=\diaght]{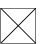}}}
\newcommand{\wigglya}{\ensuremath{\emptyset}}
\newcommand{\wigglyb}{\raisebox{\diagshift}{\includegraphics[height=\diaght]{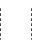}}}

\usepackage{theorem}
\theorembodyfont{\upshape}

\begin{document}


\title{Analytic calculation of $B_4$ for hard spheres in even dimensions}

\author{N.~Clisby\footnote{C.~N.~Yang Institute for Theoretical Physics, State
University of New York at Stony Brook, Stony Brook, NY 11794-3840;
e-mail: Nathan.Clisby@stonybrook.edu and
mccoy@insti.physics.sunysb.edu} \ and B.~M.~McCoy$^*$}

\maketitle

\begin{abstract}
We exactly calculate the fourth virial coefficient for hard spheres in
even dimensions for $D=4,6,8,10,$ and 12.
\end{abstract}


\section{Introduction}
\label{intro}

The virial series for the pressure 
\be \frac{P}{k_BT}=
\rho+\sum_{k=2}^{\infty}B_k \rho^{k}
\label{vir}
\ee 
of the system of hard spheres with diameter $\sigma$ in $D$ dimensions
specified by the two body
pair potential
\be
U({\bf r}) = \left\{ \begin{array}{c} +\infty \\ 0 \end{array}
\right. \begin{array}{c} |{\bf r}|<\sigma \\ |{\bf r}|> \sigma \\
\end{array}
\label{hard}
\ee 
has been studied for over 100 years.  However, despite the long
history of this problem, there are very few analytic results
known. The second virial coefficient is
\be
B_2=\frac{\sigma^D \pi^{D/2}}{ 2\, \Gamma(1+D/2)}
\ee
The third virial coefficient has been computed long ago for $D=2$
\cite{tonks1936a} and $D=3$ \cite{boltzmann1899a}, and in arbitrary dimension  
is compactly given as \cite{luban1982a}
\be
B_3/B_2^2=\frac{4\, \Gamma(1+\frac{D}{2})}{\pi^{1/2}\Gamma(\frac{1}{
2}+\frac{D}{2})}\int_0^{\pi/3} d\phi \; (\sin \phi)^D
\ee
For low dimensions these are given in \tab{virialtable}.
\renewcommand{\arraystretch}{1.2}
\begin{table*}[!htb]
\label{virialtable}
\caption{The second and third virial coefficients
as functions of dimension.}
\footnotesize
\begin{center}
\begin{tabular}{|r|l|l|l|}\hline
\multicolumn{2}{|c|}{} & \multicolumn{2}{c|}{$B_3/B_2^2$}\\
\hline
$D$&$B_2$&Exact&Numerical\\ \hline
$1$&$\sigma$&$1$&$1$\\
$2$&${\pi \sigma^2/ 2}$&$\frac{4}{3}-{\frac{\sqrt 3}{\pi}}$&$0.782004\cdots$\\
$3$&${2\pi \sigma^3/ 3}$&${5/8}$&$0.625$\\
$4$&${\pi^2 \sigma^4/ 4}$&$\frac{4}{3}-\frac{\sqrt 3}{\pi}\frac{3}{2}$&$0.506340\cdots$\\
$5$&${4\pi^2 \sigma^5/ 15}$&$53/2^7$&$0.414063\cdots$\\
$6$&${\pi^3 \sigma^6/ 12}$&$\frac{4}{3}-\frac{\sqrt 3}{\pi}\frac{9}{5}$&$0.340941\cdots$\\
$7$&${8\pi^3 \sigma^7/ 105}$&$289/2^{10}$&$0.282227\cdots$\\
$8$&${\pi^4 \sigma^8/ 48}$
&$\frac{4}{3}-\frac{\sqrt 3}{\pi}\frac{279}{140}$&$0.234614\cdots$\\
$9$&${16\pi^4 \sigma^9/945}$&$6413/2^{15}$&$0.195709\cdots$\\ 
$10$ & $\pi^5 \sigma^{10}/240$ & $\frac{4}{3}-\frac{\sqrt{3}}{\pi}\frac{297}{140}$& $0.163728\cdots$\\
$11$ & $32 \pi^5 \sigma^{11}/10395$ & $35995/2^{18}$&$0.137310\cdots$\\
$12$ & $\pi^6 \sigma^{12}/1440$ & $\frac{4}{3}-\frac{\sqrt{3}}{\pi}\frac{243}{110}$&$0.115398\cdots$\\
\hline
\end{tabular}
\end{center}
\normalsize
\end{table*}
\renewcommand{\arraystretch}{1}

However,
$B_4$ has been calculated analytically in only two and three dimensions  
\bea 
\frac{B_4}{B_2^3}&=&\left\{ \begin{array}{ll} 2-\frac{9\sqrt{3}}{2\pi}+\frac{10}{\pi^2} & {\rm for} \; D=2 { \mbox{\cite{rowlinson1964a,hemmer1964a}} }\\ 
\frac{219\sqrt{2}}{2240\pi}-\frac{89}{280}+\frac{4131}{2240\pi}
\arctan{\sqrt 2} & {\rm for} \; D=3 { \mbox{\cite{boltzmann1899a,vanlaar1899a,nijboer1952a}}} \end{array} \right.
\eea 
and no further analytic results are known. An interesting discussion
of the history of the calculation of $B_4$ in three dimensions is
given in \cite{kilpatrick1971a}.

We feel that for such a problem
in pure geometry that tractable analytic results must exist.
In this paper we make a modest step toward verifying this by
extending the analytic evaluation of the fourth virial coefficient to
even dimensions $D\leq 12.$ 

In the Mayer formalism \cite{mayer1940a} the
fourth virial coefficient is 
\be B_4 =-\frac{1}{8} \,\mayerivc-\frac{3}{4} \,
\mayerivb-\frac{3}{8} \, \mayeriva 
\label{b4mayereq}
\ee
where each solid line represents the Mayer $f$ function
\be
f({\bf r})=\exp\left({-U({\bf r})/k_BT}\right)-1
\ee
which for the hard sphere potential reduces to
\be f({\bf r}) = \left\{ \begin{array}{r} -1 \\ 0
\end{array} \right. \begin{array}{l} |{\bf r}|< \sigma \\ |{\bf
r}|>\sigma\\ \end{array}
\label{fdef}
\ee
The second and third diagrams in this expansion have been evaluated in
arbitrary dimension in \cite{luban1982a,joslin1982a}. 
In this paper we complete the computation
in even dimensions for $D\leq 12$ by evaluating the first diagram in
\sect{analyticsec}.

For virial coefficients of order greater than four it is much more
efficient to use the expansion of
Ree and Hoover \cite{ree1964a,ree1964c} where 
in addition to the $f$ bonds we also have bonds ${\tilde f}({\bf
r})=1+f({\bf r})$ which are represented by dotted lines. In this notation every
point is connected to every other point by either an $f$ or an
${\tilde f}$ bond and for $B_4$ the virial coefficient is given by
\be
B_4=\frac{1}{4}\, \wigglya - \frac{3}{8} \, \wigglyb 
\label{b4rhwigglyeq}
\ee 
where only the $\tilde f$ bonds are shown. The integral in the first
term is identical with the integral in the first term of \eqn{b4mayereq}  
and we compute the second term from the three Mayer diagrams of
\eqn{b4mayereq} in \sect{ringsec}. The results for $B_4$ and for the two
separate Ree-Hoover diagrams are given in \tab{analyticalb4table}.
\begin{table*}[bth]
\caption{Analytical results for the four point Ree-Hoover diagrams and
$B_4$ in even dimensions, with numerical values.}
\label{analyticalb4table}
\scriptsize
\begin{center}
\begin{tabular}{|r|c|c|c|}
\hline $D$ & $\wigglya/B_2^3$ & $\wigglyb/B_2^3$ & $B_4/B_2^3$ \\ \hline 
2 &$8-\frac{12\sqrt{3}}{\pi}+\frac{8}{\pi^2}$&$\frac{4\sqrt{3}}{\pi}-\frac{64}{3\pi^2}$&$2-\frac{9\sqrt{3}}{2\pi}+\frac{10}{\pi^2}$
\\ & $2.194622724\cdots$ &$0.043796999\cdots$ & $0.532231807\cdots$\\ \hline 4
&$8-\frac{18\sqrt{3}}{\pi}+\frac{238}{9\pi^2}$&$\frac{6\sqrt{3}}{\pi}-\frac{4276}{135\pi^2}$&$2-\frac{27\sqrt{3}}{4\pi}+\frac{832}{45\pi^2}$
\\ & $0.755462293\cdots$ & $0.098718698\cdots$ & $0.151846062\cdots$ \\ \hline 6
&$8-\frac{108\sqrt{3}}{5\pi}+\frac{37259}{900\pi^2}$&$\frac{36\sqrt{3}}{5\pi}-\frac{72151}{1890\pi^2}$&$2-\frac{81\sqrt{3}}{10\pi}+\frac{38848}{1575\pi^2}$
\\ & $0.285880282\cdots$ & $0.101618460\cdots$ & $0.033363148\cdots$ \\ \hline 8 &
$8-\frac{837\sqrt{3}}{35\pi}+\frac{5765723}{110250\pi^2}$&$\frac{279\sqrt{3}}{35\pi}-\frac{77417239}{1819125\pi^2}$&$2-\frac{2511\sqrt{3}}{280\pi}+\frac{17605024}{606375\pi^2}
$\\ & $0.114137690\cdots$ & $0.082912284\cdots$ & $-0.002557687\cdots$ \\ \hline 10 &
$8-\frac{891\sqrt{3}}{35\pi}+\frac{41696314}{694575\pi^2}$&$\frac{297\sqrt{3}}{35\pi}-\frac{1044625732}{22920975\pi^2}$&$2-\frac{2673\sqrt{3}}{280\pi}+\frac{49048616}{1528065\pi^2}$\\
& $0.047194685\cdots$ & $0.06069639\cdots$ & $-0.01096248\cdots$ \\ \hline 12 &
$8-\frac{1458\sqrt{3}}{55\pi}+\frac
{88060381669}{1344697200\pi^2}$&$\frac{486\sqrt{3}}{55\pi}-\frac{21249584434511}{445767121800\pi^2}$&$2-\frac{2187\sqrt{3}}{220\pi}+\frac{11565604768}{337702365\pi^2}$\\
& $0.020007319\cdots$ & $0.041792298\cdots$ & $-0.010670281\cdots$ \\ \hline
\end{tabular}
\end{center}
\normalsize
\end{table*}

\section{Analytical Calculation of the Complete Star Diagram}
\label{analyticsec}

\begin{figure*}[tbh]
\begin{center}
\begin{minipage}[t]{5cm}
\includegraphics[width=5cm]{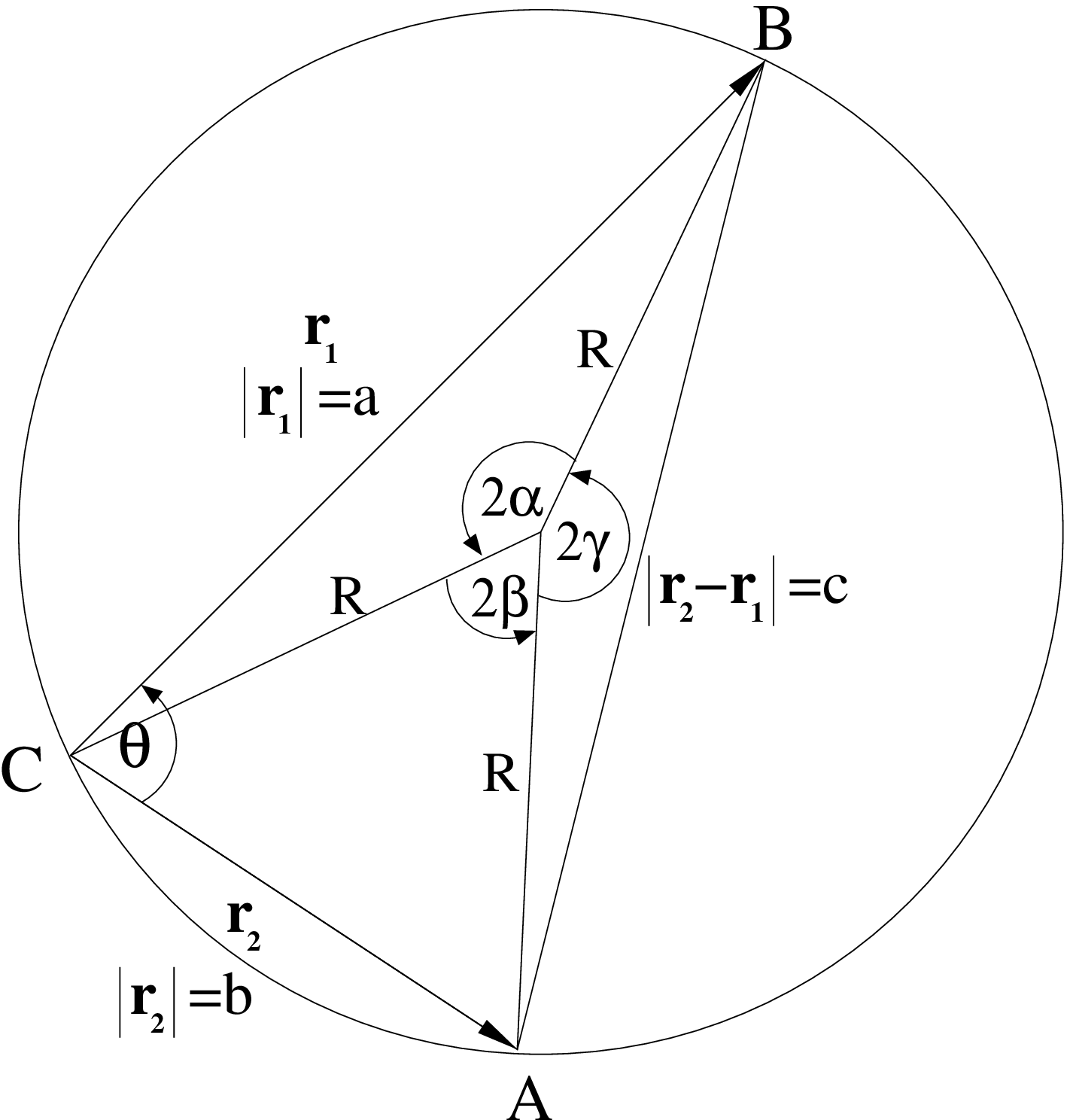}
\caption{Change of variables.}
\label{b4int}
\end{minipage}
\hspace{1cm}
\begin{minipage}[t]{5cm}
\includegraphics[width=5cm]{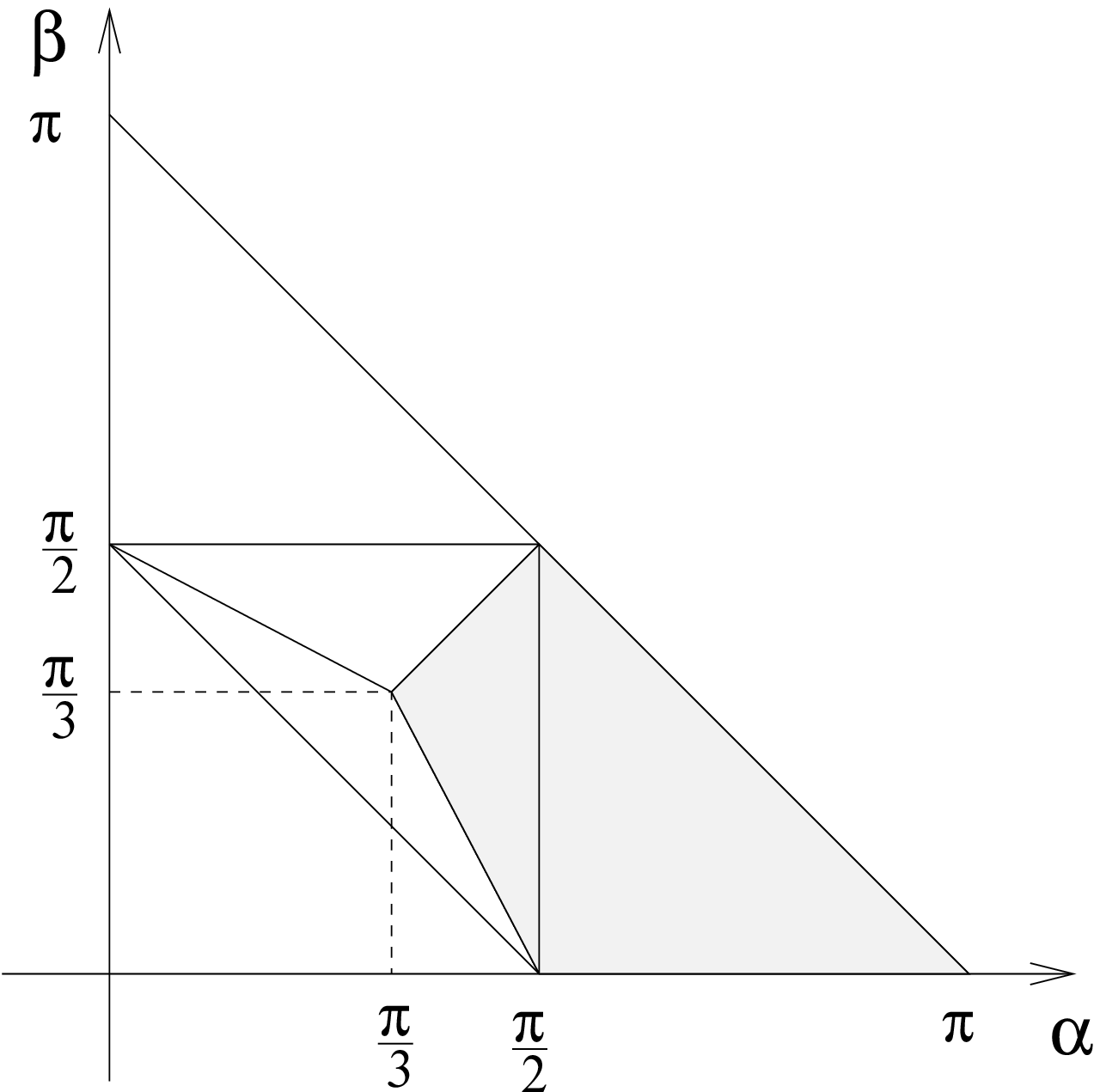}
\caption{Domain of integration for $B_4$, where the shaded region
satisfies the condition $\alpha > \beta,\gamma$.}
\label{abdomain}
\end{minipage}
\end{center}
\end{figure*}

The complete star integral in the expansions of \eqn{b4mayereq} and
\eqn{b4rhwigglyeq} is by definition
\small \begin{eqnarray} 
\mayerivc &=& \lim_{V \rightarrow \infty} \frac{1}{V} \int d^D {\bf r_1} d^D
{\bf r_2} d^D {\bf r_3} d^D {\bf r_4} f({\bf r_{12}}) f({\bf r_{13}})
f({\bf r_{14}}) f({\bf r_{23}}) f({\bf r_{24}}) f({\bf r_{34}})\nn &=&
\int d^D {\bf r_1} d^D {\bf r_2} d^D {\bf r_3} f({\bf r_{12}}) f({\bf
r_{13}}) f({\bf r_{23}}) f({\bf r_{1}}) f({\bf r_{2}}) f({\bf r_{3}})
\label{defcstar}
\end{eqnarray} \normalsize
where ${\bf r_{ij}} = {\bf r_i-r_j}.$ Specializing to the case of hard
spheres we first note that $f({\bf r})\equiv f(|{\bf r}|)$, and then
we treat the coordinates in \eqn{defcstar} as follows: ${\bf
r_1}$ is constrained to a unit ball centered on the origin due to
$f(r_1)$, ${\bf r_2}$ is integrated in the same ball
with the additional condition that $r_{12} \equiv |{\bf r_{12}}| = \sqrt{r_1^2+r_2^2-2r_1r_2\cos\theta}< 1$. The ${\bf
r_3}$ integral may be thought of as giving the overlapping hypervolume
$V_D$ of $D$--dimensional unit balls situated at the origin, ${\bf
r_1}$, and ${\bf r_2}$, and this depends only on $r_1$, $r_2$, and the
angle $\theta$ between ${\bf r_1}$ and ${\bf r_2}$. We define $V_D$ as
\be V_D(r_1,r_2,\theta) = -\int d^D{\bf r_3} f({\bf r_{13}}) f({\bf
r_{23}}) f({\bf r_{3}}) \ee where the negative sign ensures that $V_D$
is positive. We may substitute this in to the expression for
\mayerivc, and then write ${\bf r_1}$ and ${\bf r_2}$ in
$D$--dimensional spherical polar coordinates \bea \mayerivc &=& -\int
d^D {\bf r_1} d^D {\bf r_2} f(r_1) f(r_2)
f(r_{12}) V_D(r_1,r_2,\theta) \nn 
&=&-\int dr_1 r_1^{D-1} \int d\Omega_{D-1}^{(1)} \int dr_2 r_2^{D-1} \int
d\theta (\sin\theta)^{D-2} \int d\Omega_{D-2}^{(2)}
\nn && \times f(r_1) f(r_2) f(r_{12})
V_D(r_1,r_2,\theta)
\label{cstar1}
\eea
where the angular integrals give $\Omega_{D-1} \equiv \int
d\Omega_{D-1} = 2 \pi^{D/2}/\Gamma(D/2)$,
which is valid for arbitrary $D > 0$, including non-integer $D$.
The integrand in \eqn{cstar1} only depends on $r_1$,
$r_2$, and $\theta$, and we may integrate out the other angles to obtain
\bea \mayerivc &=& -\Omega_{D-1} \Omega_{D-2} \int dr_1 r_1^{D-1} \int
dr_2 r_2^{D-1} \int d\theta (\sin\theta)^{D-2}  \nn && \times f(r_1)
f(r_2) f(r_{12}) V_D(r_1,r_2,\theta)
\label{cstar2}
\eea
We now change coordinates from $(r_1,r_2,\theta)$ to the
coordinate system $(R,\alpha,\beta)$ which is illustrated in
\fig{b4int}. 
The three points which were at the origin, ${\bf r_1}$,
and ${\bf r_2}$ are now circumscribed by a circle of radius $R$;
$2\alpha$ and $2\beta$ are the angles subtended by $r_1$ and $r_2$
from the center of the circle.  \bea r_1 &=& 2 R \sin\alpha \nn r_2
&=& 2 R \sin\beta \nn \theta &=& \pi-\alpha-\beta \eea
Noting also that $-f(r_1) f(r_2) f(r_{12})$ may only have the values
of 0 and 1, we may absorb the $f$ functions in to the domain of
integration and rewrite the triple integral of \eqn{cstar2} as \small \begin{equation} 4^D
\int d\alpha \int d\beta \int dR R^{2D-1}
\left[\sin\alpha\sin\beta\sin(\alpha+\beta)\right]^{D-1}
V_D(R,\alpha,\beta)
\label{cstaralpha}
\end{equation} \normalsize

We are required to integrate over all $R$, $\alpha$, and
$\beta$ such that the distance between any two points is less than
one; denoting the sides opposite points A, B, and C on \fig{b4int} as
a, b, and c respectively we have $a = 2 R \sin\alpha$, $b = 2 R \sin\beta$,
and $c = 2 R \sin\gamma$.
We choose to integrate over $R$ before the angles $\alpha$
and $\beta$, and we impose
$b, c < a < 1$ which introduces an extra factor of $3$. Therefore we
must change the region of integration from 
\bea 0<&b<&a \label{ineq2} \\
0<&c<&a \label{ineq3} \\
0<&a<&1 \label{ineq1}
\eea
to a domain for $(R,\alpha,\beta)$. The boundary of this
region is found by setting inequalities \ref{ineq2} and \ref{ineq3} to
equality, with solutions: \bea \alpha&=&\beta \nn \alpha&=&\pi-\beta
\nn \alpha&=&\gamma \Rightarrow \beta=\pi-2\alpha \nn
\alpha&=&\pi-\gamma \Rightarrow \beta=0 \eea
This region is shown in \fig{abdomain}. Finally, inequality
\ref{ineq1} restricts $R$ to the domain 
\be 
0<R<\frac{1}{2\sin\alpha} \label{ineqR}
\ee

The next step is to calculate the overlapping volume $V_D$ of the
three hyperspheres for even dimensions $D=2n$, where $n$ is an integer
greater than or equal to 2. We do this in
Appendix \ref{analyticb4}, and the result for $R<1$ is denoted $V_D^<$
and given in \eqn{vd<eqn}, while the result for $R>1$ is denoted
$V_D^>$ and given in \eqn{vd>eqn}. Throughout this paper we will adopt the convention that sums with index $\xi$
are always over $\{\alpha,\beta,\gamma\}$.

We will sketch the steps
involved in the rest of the
calculation, which were carried out using the computer algebra platform 
MAPLE version 8 for $n=2,3,4,5,6$.
The first step is to perform the integral over $R$,
where we notice from inequality \ref{ineqR} that we may only have  $R>1$ when
$\alpha>\frac{5\pi}{6}$. Ignoring for the moment factors not
involving $R$ in \eqn{cstaralpha}, we therefore need to calculate \be
I_2=\int_0^{1/2\sin\alpha} dR \; R^{4n-1} V_D^< \quad
\alpha<\frac{5\pi}{6} \ee
and \be I_3=\int_0^1 dR \; R^{4n-1} V_D^< +
\int_1^{1/2\sin\alpha} dR \; R^{4n-1} V_D^> \quad \alpha>\frac{5\pi}{6}
\ee
One of the integrals we will need to do is 
\small \begin{eqnarray} \int dR \; R^{4n-1}\left(\arccos(R\sin\xi)-R\sin\xi \
\sqrt{1-R^2\sin^2\xi}\right) && \nn
=\frac{1}{4n}R^{4n}\left(\arccos(R\sin\xi)-R\sin\xi \
\sqrt{1-R^2\sin^2\xi}\right) && \nn +\frac{1}{2n}\int dR \; R^{4n}
\sin\xi \ \sqrt{1-R^2\sin^2\xi} &&\end{eqnarray} \normalsize
and thus the only non-trivial parts of the integral are in
the form $\int dR \; R^{2l} U^{m+\frac{1}{2}}$ with the shorthand notation
$U=1-R^2\sin^2\xi$.
We substitute \\ $R^2=[{1-(1-R^2\sin^2\xi)}]/{\sin^2\xi}$
everywhere, and this leaves us only to perform integrals such as 
\footnotesize
\bea 
\int dR \; U^{j+\frac{1}{2}} &=&\frac{\sqrt{U}}{2(j+1)}\left\{U^j+\sum_{k=0}^{j-1}\frac{(2j+1)(2j-1)\ldots(2j-2k+1)}{2^{k+1}j(j-1)\ldots
(j-k)} U^{j-k-1}\right\} \nn
&&+\frac{(2n+1)!!\arcsin(R\sin\xi)}{2^{j+1}(j+1)!\sin\xi} 
\eea
\normalsize
We may replace $\gamma$ everywhere in this
expression by $\beta$, as we are integrating over a region that is
symmetric in $\beta$ and $\gamma$. After doing this and substituting
in the limits of integration, we are left with terms involving
$[{1-\sin^2\beta/(4\sin^2\alpha)}]^{1/2}$ and
$\arcsin[\sin\beta/(2\sin\alpha)]$, along with functions of the form
$x^{-m} P(x)$ where $m$ is an integer, $P(x)$ is a polynomial, and $x$ may be either
$\sin\alpha$ or $\sin\beta$. We now make the
change of variables $x=\sin\alpha$, $y=\sin\beta$, and so the $\alpha$
and $\beta$ integrals become 
\scriptsize
\bea
 \lefteqn{\int d\alpha \int d\beta \;
\left[\sin\alpha\sin\beta\sin(\alpha+\beta)\right]^{2n-1}} \nn &=& \int dx
\; \frac{1}{\sqrt{1-x^2}} \int dy \;\frac{1}{\sqrt{1-y^2}} \; x^{2n-1}
y^{2n-1}\left[x\sqrt{1-y^2}\pm y\sqrt{1-x^2}\right]^{2n-1} \nn &=& \int dx
\; \frac{x^{2n-1}}{\sqrt{1-x^2}} \int dy \; \frac{y^{2n-1}}{\sqrt{1-y^2}}
\sum_{j=0}^{2n-1} {2n-1 \choose j}
\left[x\sqrt{1-y^2}\right]^j \left[\pm y\sqrt{1-x^2}\right]^{2n-1-j} \nn
&=& \sum_{j=0}^{n-1} {2n-1 \choose j}
\int dx \; x^{2n-1} \int dy \; y^{2n-1}
\left[\pm\frac{\left[x^2(1-y^2)\right]^j
\left[y^2(1-x^2)\right]^{n-1-j} \; y}{\sqrt{1-y^2}}\right. \nn
&&\left.+\frac{\left[y^2(1-x^2)\right]^j
    \left[x^2(1-y^2)\right]^{n-1-j}\; x}{\sqrt{1-x^2}}
\right] \non
\eea
\normalsize
The ambiguous sign is $+$ when $\alpha<\frac{\pi}{2}$ and
$-$ when $\alpha>\frac{\pi}{2}$. Under the change of coordinates 
$[1-\sin^2\beta/(4\sin^2\alpha)]^{1/2}$ becomes $[4x^2-y^2]^{1/2}/(2x)$,
and so naively it appears that we have to
compute elliptic integrals of the form $\int dz \sqrt{t^2-z^2}/{\sqrt{1-z^2}}$
and then use identities for
elliptic integrals to reduce the final result to a simple
form. However, by splitting the final integrals over x and y for
\mayerivc \ in to the pieces \small \bdm \pm\int dx \; x^{2n-1} \int \; dy
y^{2n-1}\left[x^2(1-y^2)\right]^j \left[y^2(1-x^2)\right]^{n-1-j}\frac{y}{\sqrt{1-y^2}} \; (I_2
\ \mathrm{or} \ I_3) \edm
\bdm +\int dx \; x^{2n-1} \int dy \; y^{2n-1}
\left[y^2(1-x^2)\right]^j \left[x^2(1-y^2)\right]^{n-1-j}\frac{x}{\sqrt{1-x^2}} \; (I_2 \
\mathrm{or} \ I_3) \edm \normalsize
we avoid this. The first integral may be completed
straightforwardly by first integrating over x, and then when
one substitutes in the limits of integration this eliminates any
elliptic integrals. The integral over y
may now be performed without requiring any elliptic
functions. Conversely, the second integral is completed by first
integrating over y and subsequently over x. Thus one can see that for any
even dimension all integrals that must be performed are elementary.

So we obtain the results of \tab{analyticalb4table} by carrying out this
procedure for $D=4,6,8,10,$ and $12$ using MAPLE. 

\section{Analytic derivation of the Ree-Hoover ring diagram}
\label{ringsec}

The second and third Mayer diagrams have been found in terms of
integrals over Bessel functions and hypergeometric
functions by Luban and Baram \cite{luban1982a}. The Mayer ring diagram is given by
\bdm
\mayeriva = \frac{2^{D+4}}{\pi}\frac{\Gamma(1+D)\left[\Gamma(1+\frac{D}{2})\right]^3}{\Gamma(1+\frac{3D}{2})\left[\Gamma(\frac{3}{2}+\frac{D}{2})\right]^2}
\;
  {}_3F_2\left(\frac{1}{2},1,\frac{1-D}{2};\frac{3+D}{2},\frac{3+D}{2};1\right) \edm
while the second Mayer diagram is given in \cite{luban1982a} as
\bdm
\mayerivb = -2^{D+1}D^3[\Gamma(D/2)]^2\int_0^2 dy \; y
\left[g_{D/2}(y)\right]^2 \edm
where  
\bdm
g_{\nu}(y) = \int_0^\infty dx \; x^{-\nu} \left[J_{\nu}(x)\right]^2J_{\nu-1}(xy)
\edm
Other expressions for these diagrams were given in \cite{luban1982a,joslin1982a},
but MAPLE was straightforwardly able to evaluate these expressions for
dimensions one through twelve, and these are listed \tab{ringtable}. The
second diagram of the Ree-Hoover expansion was then obtained from the
equation
\be
\wigglyb = \mayerivc + 2 \; \mayerivb + \mayeriva
\ee

\renewcommand{\arraystretch}{1.2}
\begin{table*}[!hbt]
\label{ringtable}
\caption[]{$\mayerivb/B_2^3$ and $\mayeriva/B_2^3$ in dimensions up to
  twelve.}
\begin{center}
\begin{tabular}{|r|l|l|}\hline
$D$\hspace{0.5cm}&$\mayerivb/B_2^3$&$\mayeriva/B_2^3$\\ 
\hline
$1$&$-\frac{14}{3}$&$\frac{16}{3}$\\
$2$&$-8+\frac{8\sqrt{3}}{\pi}+\frac{20}{3\pi^2}$&$8-\frac{128}{3\pi^2}$\\
$3$&$-\frac{6347}{3360}$&$\frac{272}{105}$\\
$4$&$-8+\frac{12\sqrt{3}}{\pi}+\frac{173}{135\pi^2}$&$8-\frac{8192}{135\pi^2}$\\
$5$&$-\frac{20830913}{24600576}$&$\frac{4016}{3003}$\\
$6$&$-8+\frac{72\sqrt{3}}{5\pi}-\frac{193229}{37800\pi^2}$&$8-\frac{65536}{945\pi^2}$\\
$7$&$-\frac{87059799799}{217947045888}$&$\frac{296272}{415701}$\\
$8$&$-8+\frac{558\sqrt{3}}{35\pi}-\frac{76667881}{7276500\pi^2}$&$8-\frac{134217728}{1819125\pi^2}$\\
$9$&$-\frac{332647803264707}{1711029608251392}$&$\frac{1234448}{3187041}$\\ 
$10$ &$-8+\frac{594\sqrt{3}}{35\pi}-\frac{9653909}{654885\pi^2}$&$8-\frac{1744830464}{22920975\pi^2}$\\
$11$ &$-\frac{865035021570458459}{8949618140032008192}$&$\frac{55565456}{260468169}$\\
$12$ &$-8+\frac{972\sqrt{3}}{55\pi}+\frac{182221984415}{10188962784\pi^2}$&$8-\frac{4312147165184}{55720890225\pi^2}$\\
\hline
\end{tabular}
\end{center}
\end{table*}
\renewcommand{\arraystretch}{1}

\appendix

\section{Calculation of $V_D$ in even dimensions}
\label{analyticb4}
For $D=2$ the calculation of $V_D$ is straightforward as the ${\bf r_3}$
integral is confined to the same plane as the three circle centers and
trivially gives the area of intersection of the three circles of
radius $r=1$. Although we will neglect the details for the moment, the
area depends on $R$, $\alpha$, and $\beta$ and we will denote this as
$A(r=1,R,\alpha,\beta)$.

\begin{figure*}[!tbh]
\begin{center}
\begin{minipage}[t]{5cm}
\includegraphics[width=5cm]{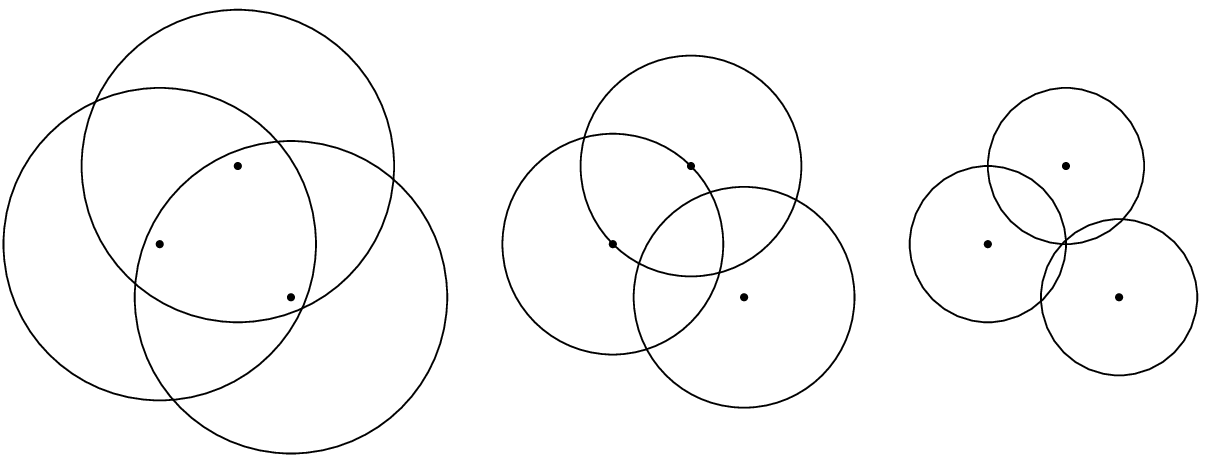}
\caption{Circles produced by cutting intersecting spheres at $h=0$,
$h={1}/{\sqrt{2}}$, and $h=\sqrt{1-R^2}={\sqrt{3}}/{2}$, where
circle centers form an acute angled triangle.}
\label{vol2}
\end{minipage}
\hspace{1.0cm}
\begin{minipage}[t]{5cm}
\includegraphics[width=5cm]{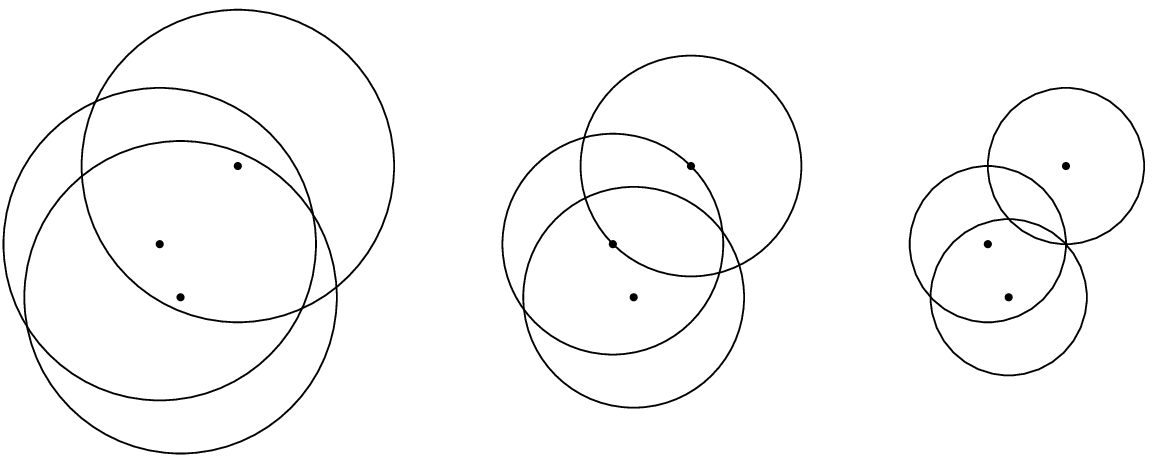}
\caption{Circles produced by cutting the intersecting spheres at
$h=0$, $h={1}/{\sqrt{2}}$, and
$h=\sqrt{1-R^2}={\sqrt{3}}/{2}$. In this case the circle centers
form a triangle with an obtuse angle and the area of intersection will
be non-zero even for $h > \sqrt{1-R^2}$.}
\label{vol3}
\end{minipage}
\end{center}
\end{figure*}

As shown in \fig{vol1}, for $D=3$, we define the
perpendicular distance from the plane of the circle centers to be $h$;
$h=0$ is the original plane in which we see three intersecting circles
of radius one. As we increase $h$ we see overlapping circles with the
same center but decreasing radius, with the radii of these circles
given by $r=\sqrt{1-h^2}$ (see Figs. \ref{vol2} and \ref{vol3}). The total
volume of intersection may be obtained by integrating the overlapping
area of the three circles with respect to $h$, from $h=0$ to the value
\hmax \ where the intersection of the three circles is reduced to a
single point. There is an additional factor of two because we need to
integrate both above and below the plane.  \bea V_{3}(R,\alpha,\beta)
&=& 2\int_0^{\hmax}dh A(r,R,\alpha,\beta)|_{r=\sqrt{1-h^2}} \nn &=&
2\int_{\rmin}^{1} dr \; \frac{r}{\sqrt{1-r^2}} A(r,R,\alpha,\beta) \eea
\setlength{\unitlength}{0.57mm}%
\begin{figure*}[!bth]
\begin{center}
\begin{picture}(100,100)(0,0)%
\put(-26,46){h} \put(-19,38){\vector(0,1){20}}%
\put(-20,0){\includegraphics[width = 8cm]{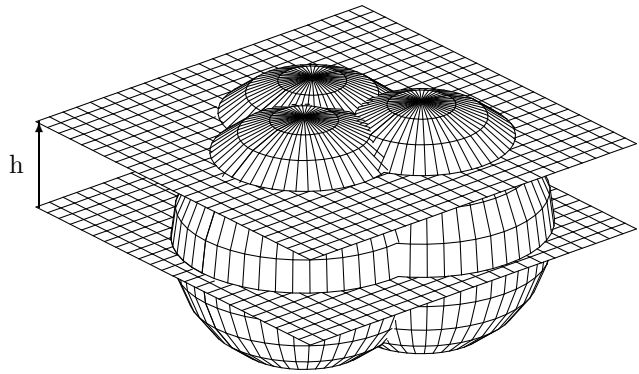}}%
\end{picture}%
\setlength{\unitlength}{1.mm}%
\end{center}
\caption{Intersecting three dimensional spheres.}
\label{vol1}
\end{figure*}

For the general case of $D > 3$, we choose a spherical polar
coordinate system for the $D-2$ dimensional subspace with radial
coordinate $h$; there is an extra factor of $h^{D-3}$ compared to three
dimensional case, and we now need to perform an additional trivial
angular integration, ie \bea V_{D}(R,\alpha,\beta)&=&\int_0^{\hmax}dh
h^{D-3} \int d\Omega_{D-3} A(r,R,\alpha,\beta)|_{r=\sqrt{1-h^2}} \nn
&=&\Omega_{D-3} \int_0^{\hmax}dh h^{D-3}
A(r,R,\alpha,\beta)|_{r=\sqrt{1-h^2}} \nn &=&\Omega_{D-3}
\int_{\rmin}^1 dr \; r (1-r^2)^{{(D-4)}/{2}} A(r,R,\alpha,\beta)
\label{dvolumeeqn}
\eea
Although \eqn{dvolumeeqn} breaks down for $D \le 2$, note
that it is correct for $D=3$ as $\Omega_0=2$.

As the area of intersection may be bounded by arcs from
either two or three circles, we will need to deal with two different
expressions for the area, and three different cases for $V_D$. In
\fig{vol1} we have $R<1$ and $\alpha<\frac{\pi}{2}$ in which case the
area is bounded by three arcs over the full range of $r$, while in
\fig{vol3} we have $R<1$ and $\alpha>\frac{\pi}{2}$ for which the area
is bounded by three arcs for $R<r<1$, but by two arcs for
$R\sin\alpha<r<R$. When $R>1$ the area can only have two arcs for its
boundary, and this is the third case.

\begin{figure*}
\begin{center}
\includegraphics[width=8cm]{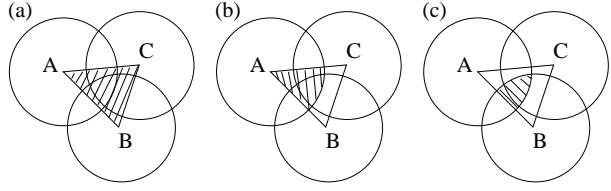}
\end{center}
\caption{The different kinds of contributions to the
inclusion-exclusion formula for the area of intersection of three
circles: (a) Area of $\bigtriangleup ABC$ (b) Area of sector A (c)
Half of the area of intersection of circles A and B.}
\label{inclexcl}
\end{figure*}

Provided the overlapping region is bounded by arcs from
three circles, the area of overlap may be calculated by an
inclusion-exclusion method following the method used for the
overlapping volume of 3 spheres in \cite{powell1964a}. The area of
intersection of three circles of radius r is given by 
\bea
 \lefteqn{A^{(3)}(r,R,\alpha,\beta)} \nn
&=& (\mathrm{Area \ of \ }
\bigtriangleup ABC) - \sum_{A_i=\{A,B,C\}}(\mathrm{Area \ of \
sector \ within \ circle \ A_i} ) \nn &&
+\frac{1}{2} \sum_{\{A_i,A_j\}} (\mathrm{Area \ of \ intersection \ of
\ circles \ A_i \ and \ A_j}) \nn &=& \frac{1}{2} R^2
(\sin2\alpha+\sin2\beta+\sin2\gamma) -
\frac{1}{2}r^2(\alpha+\beta+\gamma)\nn 
&&+\sum_{\xi}
\left(r^2\arccos\left(\frac{R
\sin\xi}{r}\right)-R\sin\xi \ \sqrt{r^2-R^2\sin^2\xi}\right) \nn
&=& R^2 \sum_{\xi} \sin\xi\cos\xi  \; - \frac{1}{2}\pi r^2 \nn
&&+\sum_{\xi} \left(r^2\arccos\left(\frac{R
\sin\xi}{r}\right)-R\sin\xi \ \sqrt{r^2-R^2\sin^2\xi} \right)
\label{inclexcleq2}
\eea
\normalsize
The different contributions to \eqn{inclexcleq2} are shown
explicitly in \fig{inclexcl}, and one should also note that although
expressions are written in terms of $\alpha$, $\beta$, and $\gamma$ for
the sake of simplicity, these variables are not independent and
$\alpha+\beta+\gamma=\pi$.
The overlap of two circles of radius r separated by a
distance $R\sin\alpha$ is given by \be A^{(2)}(r,R,\alpha) = 2
\left(r^2\arccos\left(\frac{R \sin\alpha}{r}\right)-R\sin\alpha \
\sqrt{r^2-R^2\sin^2\alpha}\ \right) \ee
We may obtain this expression from \eqn{inclexcleq2} by
taking the limit $r\rightarrow R^+$ when $\alpha>\frac{\pi}{2}$. Thus
the three expressions for $V_D$ are 
\bea
\frac{V_D}{\Omega_{D-3}} &=& \int_R^1dr \; r \, (1-r^2)^{{(D-4)}/{2}} A^{(3)} \quad \alpha<\frac{\pi}{2} \label{veq1} \\
\frac{V_D}{\Omega_{D-3}} &=& \int_R^1dr \; r \,(1-r^2)^{{(D-4)}/{2}} A^{(3)}
\nn
&&+\int_{R\sin\alpha}^R dr \; r \, (1-r^2)^{{(D-4)}/{2}}
A^{(2)} \quad \alpha>\frac{\pi}{2}\ R<1 \label{veq2} \\
\frac{V_D}{\Omega_{D-3}} &=& \int_{R\sin\alpha}^1 dr \; r \, (1-r^2)^{{(D-4)}/{2}}
A^{(2)} \quad \alpha>\frac{\pi}{2}\ R>1
\eea

We will now specialize to the case where $D=2n$, where $n$
is an integer greater than or equal to $2$. By defining
\small
\bea
I_1(r,\xi)&\equiv&\int_0^r dr \; r^3 (1-r^2)^{n-2}
\left[\arccos\left(\frac{R \sin\xi}{r}\right)-\frac{R\sin\xi}{r}
\sqrt{1-\left(\frac{R\sin\xi}{r}\right)^2}\right] \non
\eea
\normalsize
and integrating out the remaining pieces, we have for $\alpha<{\pi}/{2}$, $R<1$:

\small \bea
\frac{V_D}{\Omega_{D-3}}&=&\int_R^1 dr \; r (1-r^2)^{n-2} \left[R^2
\sum_{\xi} \sin\xi\cos\xi  \;- \frac{1}{2}\pi r^2 \right] \nn
&&+\sum_{\xi} \left(I_1(1,\xi)-I_1(R,\xi)\right)\nn 
&=&
-\frac{(1-R^2)^{n-1}}{2(n-1)}R^2 \sum_{\xi} \sin\xi\cos\xi  \;+
\left(\frac{(1-R^2)^n}{2n}-\frac{(1-R^2)^{n-1}}{2(n-1)}\right)
\frac{\pi}{2} \nn && + \sum_{\xi}
\left(I_1(1,\xi)-I_1(R,\xi)\right) \non
\eea \normalsize
while for $\alpha>{\pi}/{2}$, $R<1$:
\small \bea
 \frac{V_D}{\Omega_{D-3}}&=&
-\frac{(1-R^2)^{n-1}}{2(n-1)}R^2 \sum_{\xi}
\sin\xi\cos\xi  \; +
\left(\frac{(1-R^2)^n}{2n}-\frac{(1-R^2)^{n-1}}{2(n-1)}\right)
\frac{\pi}{2} \nn && + \sum_{\xi}
\left(I_1(1,\xi)-I_1(R,\xi)\right)
+2(I_1(R,\alpha)-I_1(R\sin\alpha,\alpha)) \nn 
&=& -\frac{(1-R^2)^{n-1}}{2(n-1)}R^2 \sum_{\xi} \sin\xi\cos\xi  \; +
\left(\frac{(1-R^2)^n}{2n}-\frac{(1-R^2)^{n-1}}{2(n-1)}\right)
\frac{\pi}{2} \nn && + \sum_{\xi}
I_1(1,\xi)+I_1(R,\alpha)-I_1(R,\beta)-I_1(R,\gamma)-2I_1(R\sin\alpha,\alpha))
\non
\eea \normalsize
and finally for $\alpha>\frac{\pi}{2}$, $R>1$:
\bea
\frac{V_D}{\Omega_{D-3}}&=&
2(I_1(1,\alpha)-I_1(R\sin\alpha,\alpha)) \non
\eea \normalsize

Integrating $I_1$ once by parts 
\small
\bea 
I_1(r,\xi)&=&\left[\frac{(1-r^2)^{n-2}}{2n}-\frac{(1-r^2)^{n-1}}{2(n-1)}+\frac{1}{2n(n-1)}\right] \nn
&&\times \left[
\arccos\left(\frac{R\sin\xi}{r}\right)-\frac{R\sin\xi}{r}\sqrt{1-\left(\frac{R\sin\xi}{r}\right)^2}\right]
\nn &&-\sum_{j=0}^{n-2} \frac{(-1)^j}{j+2} {n-2 \choose j}
\int_0^r dr \; r^{2j+1} R \sin\xi \
\sqrt{r^2-R^2 \sin^2\xi} \nn 
&=&
\left[\frac{(1-r^2)^{n-2}}{2n}-\frac{(1-r^2)^{n-1}}{2(n-1)}+\frac{1}{2n(n-1)}\right] \nn
&&\times \left[
\arccos\left(\frac{R\sin\xi}{r}\right)-\frac{R\sin\xi}{r}\sqrt{1-\left(\frac{R\sin\xi}{r}\right)^2}\right]
\nn &&-\sum_{j=0}^{n-2} \frac{(-1)^j}{j+2}{n-2 \choose j} \nn
&&\times\int_0^r dr \; r ((r^2-R^2\sin^2\xi) +
R^2\sin^2\xi)^j R \sin\xi \ \sqrt{r^2-R^2 \sin^2\xi} \nn &=&
\left[\frac{(1-r^2)^{n-2}}{2n}-\frac{(1-r^2)^{n-1}}{2(n-1)}+\frac{1}{2n(n-1)}\right] \nn
&&\times \left[
\arccos\left(\frac{R\sin\xi}{r}\right)-\frac{R\sin\xi}{r}\sqrt{1-\left(\frac{R\sin\xi}{r}\right)^2}\right]
\nn &&-\sum_{j=0}^{n-2}
\sum_{k=0}^j\frac{(-1)^j}{j+2}{n-2 \choose j}
{j \choose k}
(R\sin\xi)^{2j-2k+1} \nn
&&\times\int_0^r dr \; r
(r^2-R^2\sin^2\xi)^{k+\frac{1}{2}} \nn &=&
\left[\frac{(1-r^2)^{n-2}}{2n}-\frac{(1-r^2)^{n-1}}{2(n-1)}+\frac{1}{2n(n-1)}\right] \nn
&&\times \left[
\arccos\left(\frac{R\sin\xi}{r}\right)-\frac{R\sin\xi}{r}\sqrt{1-\left(\frac{R\sin\xi}{r}\right)^2}\right]
\nn &&-\sum_{j=0}^{n-2}
\sum_{k=0}^j\frac{(-1)^j}{j+2}{n-2 \choose j}
{j \choose k}
(R\sin\xi)^{2j-2k+1} \nn
&&\times \frac{ (r^2-R^2\sin^2\xi)^{k+\frac{3}{2}}}{2k+3}
\eea
\normalsize
where we have used
\bea
\int_0^r dr \; r^3 (1-r^2)^{n-2} &=& \int_0^r dr \; r (1- (1-r^2)) (1-r^2)^{n-2} \nn
&=& -\frac{(1-r^2)^{n-1}}{2(n-1)}
+\frac{(1-r^2)^{n}}{2n}+\frac{1}{2n(n-1)} \nn &=&
\frac{1}{2}\sum_{j=0}^{n-2} \frac{(-1)^j}{j+2}
{n-2 \choose j} r^{2j+4}
\eea
and
\bea && \frac{d}{dr} \left[
  \arccos\left(\frac{R\sin\xi}{r}\right)-\frac{R\sin\xi}{r}\sqrt{1-\left(\frac{R\sin\xi}{r}\right)^2}\right] \nn 
&=& \frac{2R\sin\xi}{r^2}\sqrt{1-\left(\frac{R\sin\xi}{r}\right)^2} \eea

Substituting in the limits of integration, $r=1$
\footnotesize \bea
I_1(1,\xi)&=&\frac{1}{2n(n-1)}\left[ \arccos(R\sin\xi)- R\sin\xi \
\sqrt{1-R^2\sin^2\xi}\right] \nn &&-\sum_{j=0}^{n-2}
\sum_{k=0}^j\frac{(-1)^j}{j+2}{n-2 \choose j}
{j \choose k}
(R\sin\xi)^{2j-2k+1} \frac{(1-R^2\sin^2\xi)^{k+\frac{3}{2}}}{2k+3} 
\eea \normalsize
Substituting $r=R$
\footnotesize \bea
I_1(R,\xi)&=&\left[\frac{(1-R^2)^{n-2}}{2n}-\frac{(1-R^2)^{n-1}}{2(n-1)}+\frac{1}{2n(n-1)}\right] \nn
&&\times\left[ \arccos(\sin\xi)-\sin\xi\sqrt{\cos^2\xi}\right] \nn
&&-\sum_{j=0}^{n-2}
\sum_{k=0}^j\frac{(-1)^j}{j+2}{n-2 \choose j}
{j \choose k}
(R\sin\xi)^{2j-2k+1} \frac{R^{2k+3}(\cos^2\xi)^{k+\frac{3}{2}}}{2k+3}
 \nn
&=&\sgn\left(\frac{\pi}{2}-\xi\right)\Biggl\{\left[\frac{(1-R^2)^{n-2}}{2n}-\frac{(1-R^2)^{n-1}}{2(n-1)}+\frac{1}{2n(n-1)}\right] \nn
&&\times\left[ \frac{\pi}{2}-\xi-\sin\xi\cos\xi\right] \nn
&&-\sum_{j=0}^{n-2}
\sum_{k=0}^j\frac{(-1)^j}{j+2}{n-2 \choose j}
{j \choose k}
R^{2j+4} (\sin\xi)^{2j-2k+1} \frac{(\cos\xi)^{2k+3}}{2k+3} \Biggr\}
\nn &=&\sgn\left(\frac{\pi}{2}-\xi\right) I^<(R,\xi)
\eea \normalsize
which defines $I^<(R,\xi)$. Substituting $r=R\sin\xi$
\bea
I_1(R\sin\xi,\xi)&=&0
\eea

Substituting $I_1(R,\xi)=\sgn(\frac{\pi}{2}-\xi) I^<(R,\xi)$ in to
$V_D$, we notice that the two cases of $\alpha<\pi/2$ and
$\alpha>{\pi}/{2}$ with $R<1$ reduce to the same form, which we
will denote as $V_D^<$. Similarly $V_D^>$ denotes the volume of
overlap when $R>1$.
\small \bea
\frac{V_D^<}{\Omega_{D-3}} &=& -\frac{(1-R^2)^{n-1}}{2(n-1)}R^2
\sum_{\xi} \sin\xi\cos\xi  \; +
\left(\frac{(1-R^2)^n}{2n}-\frac{(1-R^2)^{n-1}}{2(n-1)}\right)
\frac{\pi}{2} \nn && + \sum_{\xi}
\left(I_1(1,\xi)-I^<_1(R,\xi)\right) \quad R<1 \nn
\frac{V_D^>}{\Omega_{D-3}}&=& 2 I_1(1,\alpha) \quad R>1
\eea \normalsize

Finally, by substituting the expressions for $I_1$ and using
$\alpha+\beta+\gamma=\pi$ we obtain for $R<1$:
\small \bea
\frac{V_D^<}{\Omega_{D-3}}&=&
\frac{1}{2n(n-1)}\sum_{\xi}\left(\arccos(R\sin\xi)-R\sin\xi\
\sqrt{1-R^2\sin^2\xi}\right) \nn
&&-\sum_{\xi}\sum_{j=0}^{n-2}
\sum_{k=0}^j\frac{(-1)^j}{j+2}{n-2 \choose j}
{j \choose k}
(R\sin\xi)^{2j-2k+1} \frac{(1-R^2\sin^2\xi)^{k+\frac{3}{2}}}{2k+3}
\nn &&+\sum_{\xi}\sum_{j=0}^{n-2}
\sum_{k=0}^j\frac{(-1)^j}{j+2}{n-2 \choose j}
{j \choose k}
R^{2j+4} (\sin\xi)^{2j-2k+1} \frac{ (\cos\xi)^{2k+3}}{2k+3} \nn
&&+\left[-\frac{(1-R^2)^n}{2n(n-1)}+\frac{1}{2n(n-1)}\right]\sum_{\xi}\sin\xi\cos\xi
\; -\frac{1}{2n(n-1)}\frac{\pi}{2}
\label{vd<eqn}
\eea \normalsize
and for $R>1$:
\small \bea
\frac{V_D^>}{\Omega_{D-3}}&=&
\frac{1}{n(n-1)}\left(\arccos(R\sin\alpha)-R\sin\alpha\
\sqrt{1-R^2\sin^2\alpha}\right) \nn &&-2\sum_{j=0}^{n-2}
\sum_{k=0}^j\frac{(-1)^j}{j+2}{n-2 \choose j}
{j \choose k}
(R\sin\alpha)^{2j-2k+1} \nn
&&\times\frac{(1-R^2\sin^2\alpha)^{k+\frac{3}{2}}}{2k+3}
\label{vd>eqn}
\eea \normalsize


\noindent
{\bf Acknowledgement:} {This work was supported in part by the 
National Science Foundation under DMR-0073058.}


\end{document}